\documentclass[aps,prd,preprint,superscriptaddress,nofootinbib]{revtex4-1}
\usepackage{graphics}
\graphicspath{ {./images/} }
\usepackage{amssymb}
\usepackage{amsmath}
\usepackage{verbatim}
\usepackage[utf8]{inputenc}
\usepackage{bigints}
\usepackage[rightcaption]{sidecap}
\usepackage{wrapfig}
\usepackage[export]{adjustbox}
\usepackage{color}
\usepackage[bottom]{footmisc}
\usepackage[normalem]{ulem}

\newcommand{\cJ}{\mathcal{J}}
\newcommand{\cR}{\mathcal{R}}

\newcommand{\red}[1]{\textcolor{red}{#1}}
\renewcommand{\red}[1]{{\textcolor{black}{#1}}}
\renewcommand{\sout}[1]{}

\begin{document}


\title{Minimal Warm Inflation}


\author{Kim V. Berghaus}
\email{kbergha1@jhu.edu}
\affiliation{Department of Physics $\&$ Astronomy,\\The Johns Hopkins University, Baltimore, MD 21218, USA}
\author{Peter W.~Graham}
\affiliation{Stanford Institute for Theoretical Physics, Department of Physics, Stanford University, Stanford, CA 94305-4060, USA}
\author{David E. Kaplan}
\affiliation{Department of Physics $\&$ Astronomy,\\The Johns Hopkins University, Baltimore, MD 21218, USA}



\date{\today}

\begin{abstract}
Slow-roll inflation is a successful paradigm.
However we find that even a small coupling of the inflaton to other light fields can dramatically alter the dynamics and predictions of inflation.  As an example, the inflaton can generically have an axion-like coupling to gauge bosons.
Even relatively small couplings will automatically induce a thermal bath during inflation.
The thermal friction from this bath can easily be stronger than Hubble friction, significantly altering the usual predictions of any particular inflaton potential.  Thermal effects suppress the tensor-to-scalar ratio $r$ significantly, and predict unique non-gaussianities.
This axion-like coupling provides a minimal model of warm inflation which avoids the usual problem of thermal backreaction on the inflaton potential.  As a specific example, we find that hybrid inflation with this axion-like coupling can easily fit the current cosmological data.

\end{abstract}

\pacs{}

\maketitle

\tableofcontents

\section{Introduction}


The idea of an early period of cosmic inflation is a simple way to explain the near homogeneity and isotropy of the universe.
Many of the simplest single-field models are already constrained by measurements of the scalar to tensor ratio $r$ \cite{Akrami:2018odb, Martin:2013tda, Tsujikawa:2013ila}.
Warm inflation offers an interesting alternative \cite{Berera:1995ie, Berera:1995wh, Berera:1996nv, Berera:1999ws} (for review, see \cite{Rangarajan:2018tte}). It turns out to be possible to have a concurrent quasi-thermal radiation bath if energy is extracted from the rolling scalar field via friction. The benefits of warm inflation include automatic reheating at the end of inflation when the thermal bath begins to dominate over the vacuum energy and suppressing contributions to the scalar-tensor ratio $r$ \cite{Berera:1999ws, BasteroGil:2009ec}. It further enhances non-gaussianities and predicts a unique shape for the bispectrum, which is a `smoking gun' for warm inflation, making it distinguishable from all other inflationary models \cite{Bastero-Gil:2014raa}. Despite these benefits, in practice it has been challenging to embed warm inflation consistently within a microphysical theory due to large thermal backreactions on the inflaton potential \cite{Yokoyama:1998ju}, although progress has been made over the last twenty five years \cite{Berera:1999ws, Moss:2006gt, Bastero-Gil:2016qru, Bastero-Gil:2019gao}.
 
In this paper, we show that giving the inflaton an axion-like coupling naturally leads to warm inflation. This generates a thermal bath self-consistently without significant back-reaction on the inflaton potential.  The coupling can produce a simple theory of warm inflation consistent with all experimental data.  We call this Minimal Warm Inflation. 

Non-Abelian axion-like couplings in warm inflation have been considered before \cite{Visinelli:2011jy, Kamali:2019ppi}, without the explicit temperature dependence of the friction coefficient. Here, we use recent results of the sphaleron rate in classical lattice gauge theory, which predicts a dependence $\Upsilon \sim \alpha^5 \frac{T^3}{f^2}$ \cite{Moore:2010jd}. The temperature dependence greatly impacts predictions of cosmological observables \cite{Visinelli:2016rhn} such as non-gaussianities, curvature power spectrum and spectral index, and thus needs to be included.

A different class of dissipative inflationary models with axion-like couplings exist that exploit rapid gauge field production through tachyonic instabilities \cite{Anber:2009ua, Anber:2012du}. Thermalization in these models is non-trivial but can happen, leading to an alternative setup of warm inflation \cite{Ferreira:2017wlx, Ferreira:2017lnd}. In these works it has already been pointed out that the shift-symmetry of the axion can avoid thermal back-reactions.

This paper is laid out as follows:  in Section \ref{Background}, we review the general properties of inflation when it is coupled to a thermal bath and point out that warm inflation is an attractor solution.  In Section \ref{AWI}, we describe the specific case of a rolling field with an axion-like coupling to non-Abelian gauge fields and use the predicted temperature dependence to compute the power spectrum's tilt.  In Section \ref{Hybrid}, we present a specific example of a potential, that of hybrid inflation, which matches cosmological data when the axion-like coupling is included.  We present our conclusions in Section \ref{Conclusions}.  We use  Appendix \ref{sec: Weak regime appendix} to describe the part of parameter space where thermal friction is sub-dominant (so-called weak warm inflation), which could in principle allow other potentials to reproduce the data, but in a regime where the calculations of the power spectrum from thermal fluctuations have not been done explicitly.

\section{Background on Warm Inflation}\label{Background}

We now give a terse summary of warm inflation (in the strong regime) including our definitions of the slow-roll parameters and a derivation of the power spectrum.

\subsection{Framework of Warm Inflation}
We will now show that it is possible to have a quasi-steady state cosmological solution with approximately constant vacuum energy and a non-negligible thermal bath with approximately constant temperature.  We begin by considering the equation of motion of the inflaton in the presence of a temperature-dependent friction $\Upsilon$. We define the dimensionless parameter $Q \equiv \frac{\Upsilon}{3H}$, such that: 
\begin{equation} \label{EOMQ}
\ddot{\phi} + 3H \left(1+Q \right) \dot{\phi}+V'(\phi) = 0
\end{equation}
which, together with the Friedman equation, governs the inflationary dynamics:
\begin{equation} \label{friedman}
H^2 = \frac{1}{3M^2_{\text{Pl}}} \left(V(\phi) +\frac{1}{2}{\dot{\phi}}^2 +\rho_R  \right)  
\end{equation}
Here dots denote derivatives with respect to time ($\dot{\phi} = \frac{d\phi}{dt}$) and primes denote derivatives with respect to $\phi$ ($V'(\phi) =\frac{dV}{d\phi}$). During inflation the potential energy $V(\phi)$ dominates over both the kinetic energy $\frac{1}{2}{\dot{\phi}}^2$ and the radiation energy density $\rho_R$. We will see that $\rho_R$ does not decrease during slow-roll and the end of inflation can occur when $V \sim \rho_R$. A small slow-roll parameter $\epsilon_H$ ensures that the evolution of the Hubble parameter is slow with respect to time: 
\begin{equation}
\epsilon_H \equiv - \frac{\dot{H}}{H^2}   
\end{equation}
In order for accelerated expansion to be sustained, we impose another small slow-roll parameter  $\eta_H$, which we take to be: 
\begin{equation}
\eta_H \equiv  -\frac{\ddot{H}}{\dot{H}H} +\frac{\ddot{\phi}}{H \dot{\phi}} 
\end{equation}
Here we have defined $\eta_H$ such that it is independent of $\dot{Q}$. In the slow-roll regime where $\epsilon_H,\eta_H \ll 1$ we have: 
\begin{equation} \label{papprox}
\dot{\phi} \approx -\frac{V'}{3H(1+Q)}    
\end{equation}
\begin{equation} \label{Hprox}
H^2 \approx \frac{1}{3 M^2_{\text{Pl}}} V   
\end{equation}
By differentiating equations \eqref{papprox} and \eqref{Hprox} with respect to time we obtain the slow-roll parameters in terms of the potential $V(\phi)$.  To be consistent with the warm-inflation literature, we define: 
\begin{equation}
\epsilon_V \equiv   \frac{M_{\text{Pl}}^2}{2(1+Q)}\left(\frac{V'}{V}\right)^2 \simeq \epsilon_H \ll 1
\end{equation}
\begin{equation}
 \eta_V \equiv  \frac{M_{\text{Pl}}^2}{(1+Q)} \frac{V''}{V} \simeq \eta_H + \epsilon_H \ll 1
\end{equation}
Compared to cold inflation we can see that the conditions for slow-roll are relaxed due to the additional friction which permits slow-roll on steeper potentials. Thus, an advantage of warm inflation is that $\phi$ does not have to travel as far in field space to get the same number of e-folds. When $Q$ is small this is only a small suppression; however when $Q$ is large this allows sub-Planckian field values for $\phi$, while still achieving the minimally required number of e-folds, $N_{\text{CMB}} \sim 60$: 
\begin{equation} \label{NCMB}
N_{\text{CMB}}=\int H dt = \int^{\phi_{\text{CMB}}}_{\phi_{\text{end}}}\frac{1}{M^2_{\text{Pl}}} \frac{V}{V'} \left(1+Q(\phi)\right)d\phi
\end{equation}
In equation \eqref{NCMB} $\phi_{\text{CMB}}$ denotes the field value of $\phi$ at the beginning of the observable e-folds in the cosmic microwave background (CMB). $\phi_{\text{end}}$ denotes the field value of $\phi$ at the end of inflation when the universe transitions into being radiation dominated. The energy extracted from the rolling field due to the friction sources the radiation bath \cite{Mishra:2011vh}:
\begin{equation}
\label{eqn: temperature EOM}
\dot{\rho}_R + 4 H\rho_R = \Upsilon(T) \dot{\phi}^2    
\end{equation}
In the slow-roll regime where $\epsilon_V$, $\eta_V \ll 1$, we can neglect $\dot{\rho}_R$ and we obtain:
\begin{equation} \label{SS}
4 H\rho_R \approx \Upsilon(T) \dot{\phi}^2   
\end{equation}
for the quasi steady-state system. 
\\

\subsection{Predictions of Warm Inflation}
Here we focus on predictions in the strong regime ($Q \gg 1$) of warm inflation with a friction $\Upsilon \propto T^3$, which is the relevant friciton for our minimal warm inflation model as described in more detail in Section \ref{AWI}. In this regime the thermal inflaton perturbations dominate over the usually considered quantum fluctuations, as outlined in detail in, for example, \cite{Graham:2009bf}. The temperature dependence of the friction further couples the evolution of the inflaton and radiation fluctuations. This effect gives rise to a `growing mode' for the curvature power spectrum, which is absent for a temperature-independent friction coefficient or in the weak regime. The curvature power spectrum in presence of the growing mode has been calculated in \cite{Graham:2009bf}\footnote{A different calculation from \cite{BasteroGil:2011xd} suggests a scaling of $Q^{\frac{16}{2}}$ instead of $Q^{\frac{19}{2}}$. We thank Gauraw Gosmani for pointing this out. We do not take a position on this discrepancy, but note the impact on the results derived in this paper is negligible.}  for $Q \gg 1$: 
\begin{equation} \label{spectrum}
\Delta^2_R \approx \frac{\sqrt{3}}{4\pi^{\frac{3}{2}}} \frac{H^3 T}{\dot{\phi}^2}\left(\frac{Q}{Q_3}\right)^{9} Q^{\frac{1}{2}}
\end{equation}
Here $Q_3 \approx 7.3$ and is fixed by matching the boundary conditions for the solution of the inflaton perturbations in different regimes.\footnote{Approximation \eqref{spectrum} is most accurate when $Q\gg Q_3$. Reference \cite{Graham:2009bf} also provides numerical results which approximate the spectrum down to $Q = 100$. Using the more accurate numerical results makes an negligible impact on the phenomenology discussed in this paper. Thus, for easier readability we use the analytical approximation in equation \eqref{spectrum}.} 

Assuming temperatures well below the Planck scale the tensor perturbations are not affected and remain the same as the prediction for cold inflation \cite{Liddle:2000cg}:
\begin{equation} \label{ts}
\Delta_h^2 = \frac{2}{\pi^2}\frac{H^2}{{M_{\text{Pl}}^2}}
\end{equation}
The scalar to tensor ratio $r$ based on equation \eqref{spectrum} and \eqref{ts} is then given by:
\begin{equation} \label{sttr}
r \approx \frac{1}{\sqrt{3 \pi}} \frac{16 \epsilon_V}{Q^{\frac{3}{2}}} \frac{H}{T} \left(\frac{Q_3}{Q}\right)^{9}
\end{equation} 
Equation \eqref{sttr} shows that $r$ is heavily suppressed as: $\frac{H}{T} < 1$, $\epsilon_V \ll 1$, $Q \gg 1$ and $Q \gg Q_3$. This is in agreement with observational constraints as tensor modes have not been detected, yet. Contrarily, the detection of sizeable tensor modes in the future would rule out our model in the strong regime ($Q \gg 1$).

Sizeable non-gaussianities are the most distinct prediction of our minimal model of warm inflation since the total size of $f_{\text{NL}}^{\text{warm}}$ does not depend on slow-roll parameters. The strong regime of warm inflation $Q \gg 1$ has a unique dominant bispectrum shape \cite{Moss:2011qc, Bastero-Gil:2014raa}, which has been classified and constrained as 'WarmS' by the Planck 2015 results \cite{Ade:2015ava}. However, the $Q$-dependent result of $f_{\text{NL}}^{\text{warm}}$ \cite{Moss:2007cv} used in the Planck 2015 results to derive constraints on $Q$ is only valid in the absence of a temperature-dependent friction coefficient and further suffers from a sign error as was pointed out by the authors of \cite{Moss:2007cv} in subsequent work \cite{Moss:2011qc}. Considering the temperature dependence of the friction term of our 'minimal warm inflation model' gives a Q-independent prediction \cite{Bastero-Gil:2014raa}:
\begin{equation}
f_{\text{NL}}^{\text{warm}} \approx 5   
\end{equation}
This $f_{\text{NL}}^{\text{warm}}$ can be decomposed into contributions from different bispectral template shapes where $f_{\text{NL}}^{\text{warmS}} \approx 3.5$,  $f_{\text{NL}}^{\text{local}} \approx 0.5$ and $f_{\text{NL}}^{\text{equi}} \approx 1$ \cite{Bastero-Gil:2014raa}. Since the shape correlations between the 'WarmS' (equilateral) bispectral shape and the local bispectral shape is $0.27$ ($0.46$) \cite{Fergusson:2008ra}, the expected net contribution to the most constrained bispectral shape is $f_{\text{NL}}^{\text{local}} \approx 1.5$. The current most up to date constraints from Planck data are  $f_{\text{NL}}^{\text{local}} = 0.8  \pm 5$ \cite{Ade:2015ava}, which is insufficient for making conclusions about the viability of our model. While the not yet published Planck 2018 analysis may improve these bounds slightly, ideally an improvement of about a factor of $\sim 10$ in precision is needed to first discover sizeable non-gaussianities and second determine the bispectral shape. CMB Stage-4 \cite{Abazajian:2016yjj} in accordance with upcoming optical, infrared and radio surveys conducted by new experiments such as Euclid \cite{Amendola2018}, SPHEREx \cite{Dore:2018kgp}, and the SKA telescope \cite{Bacon:2018dui} respectively report possible improvements over the current errorbars by up to a factor of $10-20$ \cite{Yamauchi:2014ioa, Karagiannis:2018jdt}. Euclid (SPHEREx) is planned to be launched before 2022 (2023) whereas the construction of the first SKA telescope (SKA1) is anticipated to start at the end of 2019. If the obtained experimental data will be able to match the precision level of the forecasts we will be able to conclusively detect the level of local non-gaussianity predicted by this model, which in a subsequent analysis could potentially be distinctively attributed to warm inflation due to its unique bispectral shape \cite{Bastero-Gil:2014raa}.

\subsection{Initial Conditions for Warm Inflation}
In this subsection we show that we do not have to start with a thermal bath to achieve warm inflation.  In fact, for an inflaton that couples to light degrees of freedom with a wide range of couplings, a thermal bath will be automatically generated rapidly even starting from standard Hubble fluctuations.

If the universe starts with a low temperature it will start to heat up from the thermal friction which removes kinetic energy from the inflaton and dumps it into the thermal bath.  It will tend towards the equilibrium temperature that comes from solving eqn.~\eqref{SS}, but we want to make sure this rate is fast enough that the equilibrium temperature is reached in a short time.
To determine the time, we define constants $A$ and $B$ so that the radiation density $\rho_R = A T^4$ and the friction rate is $\Upsilon(T) = B T^p$ where we will assume the power $p < 4$ (which is the case for axion thermal friction as we will see below). We can see from eqn.~\eqref{eqn: temperature EOM} that if we start with a very low temperature then the Hubble term can be neglected and the evolution of the temperature is given by \begin{equation}
\label{eqn: growth of T}
\dot{\rho}_R \approx \Upsilon(T) \dot{\phi}^2
\end{equation}
 Then eqn.~\eqref{eqn: growth of T} gives
\begin{equation}
\label{eqn: growth of T part2}
T^{3-p} \, \frac{dT}{dt} = \frac{B \dot{\phi}^2}{4 A}
\end{equation}
we want to know that the equilibrium temperature can be reached quickly.  For this it will be enough to find an upper bound on the time required $t_\text{eq}$ to reach equilibrium.  The temperature grows faster the larger $\dot{\phi}^2$.  And note that initially at low temperatures the friction $\Upsilon(T)$ is lower than in equilibrium so the kinetic energy $\dot{\phi}^2$ will be larger (we assume here that the field $\phi$ has had time to come near its terminal velocity, but this takes at most a few e-folds).  So to find an upper bound on $t_\text{eq}$ it is conservative to assume $\dot{\phi}^2$ is fixed at its equilibrium value $\dot{\phi}_\text{eq}^2$.  Then we can solve eqn.~\eqref{eqn: growth of T part2} to find
\begin{equation}
\label{eqn: T eom soln}
    T_\text{eq}^{4-n} - T_i^{4-n} > (4-n) \frac{B \dot{\phi}_\text{eq}^2}{4 A} \, t_\text{eq}
\end{equation}
where $T_i$ is the initial temperature.
Note that the time it takes to heat up to the equilibrium temperature is essentially independent of the initial temperature (so long as it is relatively small).  This surprising fact means we can start with  any initial temperature (even quantum fluctuations of the fields would do it) and it will reach the equilibrium temperature in this same time.

In equilibrium we can solve eqn.~\eqref{SS} to find
\begin{equation}
    H T_\text{eq}^{4-n} = \frac{B \dot{\phi}_\text{eq}^2}{4 A}
\end{equation}
Putting this into eqn.~\eqref{eqn: T eom soln} we find that the time required to reach equilibrium is at most
\begin{equation}
    t_\text{eq} < \frac{1}{4-n} \frac{1}{H}
\end{equation}
So it takes less than one Hubble time to reach the equilibrium temperature for warm inflation, no matter how low the initial temperature was (even including zero initial temperature since there are always quantum mechanical fluctuations).

Note that if we start with an initial temperature in the universe which is significantly above the equilibrium temperature $T_\text{eq}$ then the temperature will drop through the normal redshifting (the Hubble term in equation \eqref{eqn: temperature EOM}).  This is not as fast as the rate we just found for the temperature approaching equilibrium from below which had the interesting behavior that it was independent of the initial temperature.  In the case of the temperature dropping towards equilibrium, it does take more than one Hubble time, but since the reshifting is exponential it only takes $\sim \ln \left( \frac{T_i}{T_\text{eq}} \right)$ e-folds of inflation before the temperature has dropped to equilibrium.

We have seen that our warm inflation is an attractor solution.  Given a potential for an inflaton, and some terms that allow the inflaton to interact with other light degrees of freedom, a thermal bath will be generated very rapidly at the start of inflation.  So it is generic to be in warm inflation instead of cold inflation, as long as the light degrees of freedom are lighter than the equilibrium temperature.  Of course if the equilibrium temperature is low enough that 
$T_\text{eq} < H$ then having this thermal bath is meaningless and we are actually in cold inflation.



\subsection{The Problems of Warm Inflation}
It is challenging to build a microphysical model that supports warm inflaton because the friction $\Upsilon$ is usually accompanied by a large thermal back-reaction onto the inflaton potential, that spoils the flatness of the potential and does not support enough e-folds. When the friction arises from perturbative interactions directly between the scalar field and light fields, the mass of the scalar fields obtains a finite temperature contributions which scales with the temperature:
\begin{equation}
\delta m^2_{\phi} \propto T^2    
\end{equation}
This correction is dominant to the friction which scales with temperature fluctuations $\Upsilon \propto \delta T$. It is possible to protect the mass of the inflaton from thermal contributions by imposing symmetries; however this generically also turns off the friction. Thus, it appears challenging to produce a large friction without unwanted mass corrections or fine-tuned cancellations.

\section{Warm Inflation with an Axion} \label{AWI}




We find a minimal warm inflation model in which the inflaton $\phi$ is an axion coupling to a pure Yang-Mills gauge group:
\begin{equation} \label{int}
\mathcal{L}_{\text{int}} = \frac{\alpha}{16 \pi}\frac{\phi}{f}\tilde{G}^{\mu\nu}_a G^a_{\mu\nu} 
\end{equation}
Here $G^a_{\mu\nu}$ ($\tilde{G}^a_{\mu\nu} = \epsilon^{\mu\nu\alpha\beta} G^a_{\alpha \beta}$) is the field strength of an arbitrary Yang-Mills group and $\alpha \equiv \frac{g_{\text{YM}}^2}{4\pi}$, and $g_{\text{YM}}$ is the gauge coupling. There is no perturbative back-reaction that scales with the temperature because the axion is protected by its shift symmetry\footnote{We softly break this symmetry by giving the inflaton a UV-potential. We have checked that the back-reaction from this breaking term is negligible.}. The back-reaction due to non-perturbative effects is just the usual axion mass, which at zero temperature scales as $\propto \frac{T^4_c}{f^2}$ and at high temperatures ($T \gg T_c$) this small quantity becomes even further surpessed as instanton methods \cite{RevModPhys.53.43} estimate a power-law decrease with $m^2_a \propto T^{-X}$, with $X \sim 7$ for pure Yang-Mills SU(3) \cite{Frison:2016vuc}, which is in agreement with lattice calculations.  This is why the back-reaction in our model is negligible.

However, at high temperatures classical transitions between vacua with different topological charge are no longer suppressed, which give rise to topological charge fluctuations. Thus, the fluctuations responsible for the friction experienced by $\phi$ are not inherently thermal; they are topological. However, the topological fluctuations still increase with temperature as higher temperatures enhance the transition rate, also known as the sphaleron rate $\Gamma_{\text{sphal}} = \underset{V,t \to \infty}{\text{lim}} \frac{\langle Q^2 \rangle}{V t}$ \cite{PhysRevD.36.581}. The friction arising from the interaction in \eqref{int} can be determined by the sphaleron rate $\Gamma_{\text{sphal}}$ in the limit of the inflaton mass being smaller than $\sim \alpha^2 T$ \cite{Laine:2016hma}:
\begin{equation}
\Upsilon(T) = \frac{\Gamma_{\text{sphal}}(T)}{2f^2T}    
\end{equation} 
The sphaleron rate has been measured within classical lattice gauge theory for pure SU(2) and SU(3) theories and indicates a scaling of $\Gamma_{\text{sphal}} \sim \alpha^5 T^4$ \cite{Moore:2010jd, Laine:2016hma}. The friction coefficient $\Upsilon$ then scales roughly as $T^3$ \cite{Moore:2010jd}:
\begin{equation} \label{friction}
\Upsilon(T) = \kappa(\alpha,N_c,N_f) \alpha^5 \frac{T^3}{f^2} 
\end{equation}
where $T$ is the temperature of the thermal bath of the Yang-Mills group and this formula only applies when that group is in thermal equilibrium\footnote{We are ignoring the weak $T$-dependence in the running of $\alpha$ as $T$ remains nearly constant during the period of inflation, and thus $\alpha$ can be treated as a fixed parameter of the model.}. $\kappa$ is an O(100) number which has a weak logarithmic dependence on $\alpha$ and whose exact value depends on the number of colors $N_c$ and flavors $N_f$ of the group \cite{Moore:2010jd}. The estimate of the friction coefficient in terms of the sphaleron rate breaks down in the weak regime of warm inflation ($Q \lesssim 1$) due to the limit $m_{\phi} \ll \alpha^2 T$ becoming oversaturated. While the mechanism itself should also work for the weak regime, we focus on the strong regime in this paper since we know the exact friction in this regime. Thermalization of the inflaton occurs in this regime if the gauge boson-inflaton scattering rate, $\Gamma_{g\phi} \approx \alpha^3 \frac{T^3}{32 \pi f^2}$ \cite{Masso:2002np, Graf:2010tv}, is much larger than the Hubble rate. This gives the  condition $\frac{3Q}{32 \pi \kappa \alpha^2} \gg 1$, which is always satisfied in the strong regime of our model, where we consider $Q > 100$ and $\alpha <0.1$.

We give the inflaton a UV-potential $V(\phi)$ (in addition to the IR potential it would get from the confining group).  We cannot use the IR potential because, in order to have a thermal bath of gauge bosons, we must have the temperature above the confinement scale.  At such temperatures the IR potential is rapidly suppressed and we have checked that it is not possible to use that potential for inflation.  So inflation occurs as the inflaton rolls down its UV potential $V$ and its equation of motion is given by:
\begin{equation} \label{EOM}
\ddot{\phi} +\left(3H + \Upsilon\right)\dot{\phi}+V'(\phi) = 0
\end{equation}

Based on the curvature power spectrum in equation \eqref{spectrum} we derive the spectral index: 
\begin{equation}
n_s-1 = \frac{d \ln\Delta_\cR^2}{dN} 
\end{equation}
\begin{equation} \label{si}
\frac{d \ln\Delta_\cR^2}{dN} =\left(\frac{5}{2}-9\right)\frac{d \ln H}{dN}-2\frac{d \ln\dot{\phi}}{dN} +\left(\frac{1}{2}+9\right) \frac{d \ln \Upsilon}{dN} + \frac{d \ln T}{dN}
\end{equation}
Using $H dt =dN$ we rewrite the derivatives in equation \eqref{si} in terms of the slow-roll parameters \cite{Ramos:2013nsa}:
\begin{equation} \label{dlnH}
\frac{d \ln{H}}{dN} = -\epsilon_V 
\end{equation}
\begin{equation} \label{dlnphid}
\frac{d {\ln \dot{\phi}}}{dN} = \epsilon_V - \eta_V -\frac{Q}{1+Q} \frac{d {\ln Q}}{dN}   
\end{equation}
\begin{equation} \label{dlnQ}
\frac{d {\ln Q}}{dN} =  \epsilon_V + 3 \frac{d {\ln T}}{dN}  
\end{equation}
We use equation \eqref{SS} to express the temperature as a function of time resulting in: 
\begin{equation} \label{dlnT}
\lim_{Q \gg 1}\frac{d {\ln T}}{dN}   =  \frac{1}{7}\left(\epsilon_V - 2 \eta_V \right)
\end{equation}
\begin{equation} \label{dlnU}
\frac{d {\ln \Upsilon}}{dN} = 3  \frac{d {\ln T}}{dN}
\end{equation}
Plugging in \eqref{dlnH}, \eqref{dlnphid}, \eqref{dlnT} and \eqref{dlnU} into \eqref{si} we find the spectral index in leading order in $\epsilon_V$ and $\eta_V$ in the strong regime of warm inflation:   
\begin{equation} \label{ns}
n_s-1=\frac{3}{7}\left(27\epsilon_V-19\eta_V\right)
\end{equation}
Compared to the spectral tilt obtained from the standard cold inflation power spectrum the sign of $\epsilon_V$ and $\eta_V$ is inverted for the strong regime of warm inflation. This conveys interesting constraints on possible potential shapes for warm inflation that are in agreement with the observed red tilt ($n_s -1< 0$), as $\eta_V$ has to be larger than $\epsilon_V$.

\section{An Example: Hybrid Warm Inflation}\label{Hybrid}

\subsection{Inflation} \label{Iitm}

\begin{figure}
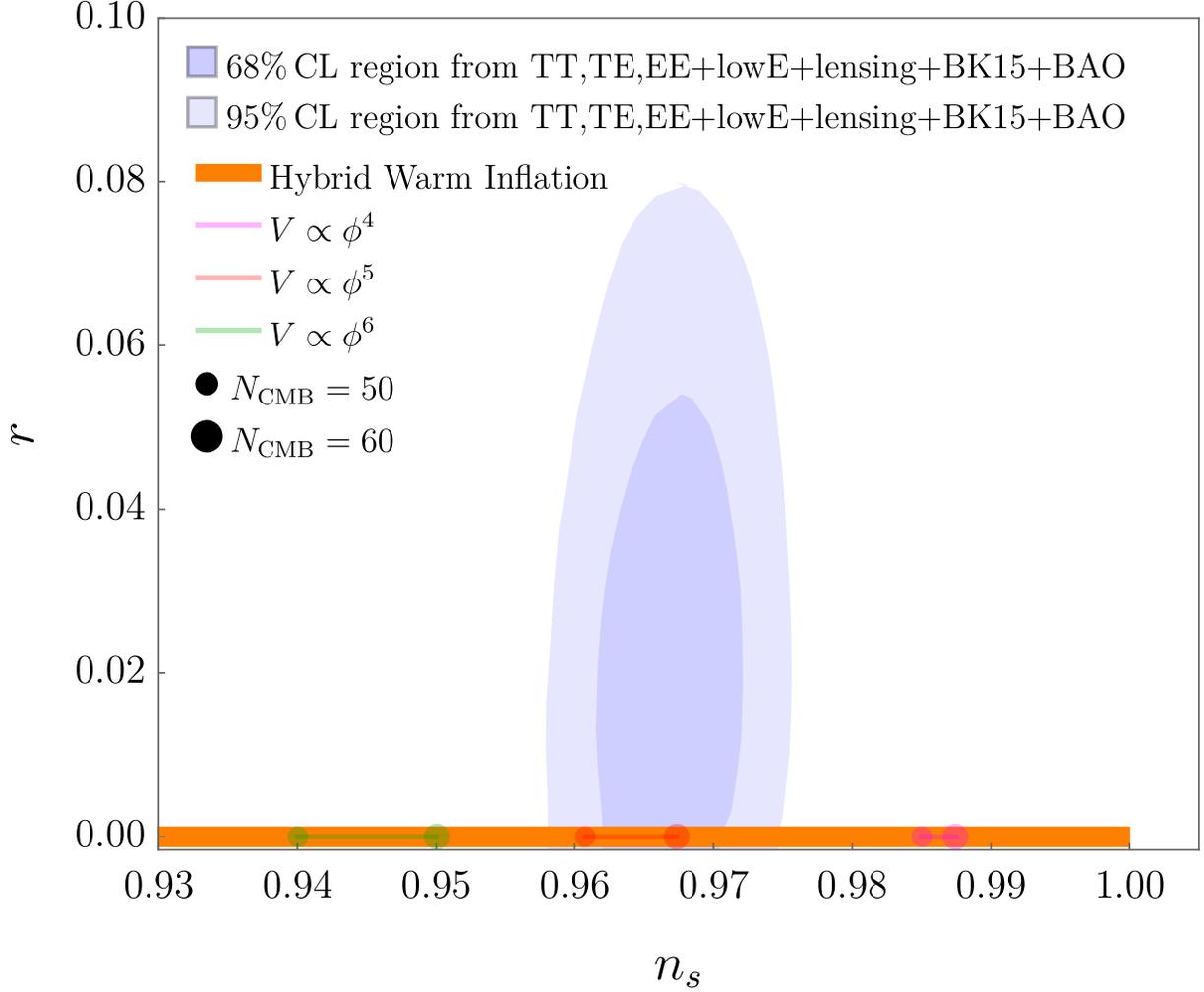
 
    \adjustimage{width=\textwidth,center}{"hybridplot2".pdf}
    \caption{Comparison of the predicted spectral index $n_s$ in the strong regime of minimal warm inflation, given different potentials.  Hybrid warm inflation overlaps with the allowed region. Due to remaining free parameters in hybrid inflation it is able to reproduce various red-tilted values of $n_s$. In single field inflation $V \propto \phi^5$ lies in the allowed region in the $r$-$n_s$ plane (although such a potential in general is not compelling as it requires extreme fine-tuning). All predictions for the tensor-to scalar ratio in the strong regime are $r \approx 0$ due to the heavy surpression of $r$ for $ Q \gg 1$. The shown allowed contour regions are the most stringent to date using Planck 2018 data as well as lensing, polarizations data from BICEP2/Keck Array BK15 and baryon acoustic oscillation (BAO).   }
    \label{rvsns}
\end{figure}

In the strong regime of warm inflation the expression for the spectral tilt in \eqref{ns} only reproduces the experimentally observed red tilt when $\epsilon_V < \eta_V$. For a single scalar field model this requires a fine-tuned level of convexity of the potential $V \propto \phi^n$ with $n \gtrsim 4$. Similarly, the lowest order cosine-like potential that is able to reproduce the observables requires $V \propto  (1 + \cos\frac{\phi}{f_{\phi}})^n$ with $n \geq 3$. In particular, a single cosine does not fit the observations. As an example, Figure~\ref{rvsns} shows how $V \propto \phi^5$ can reproduce the observed spectral index in single field inflation. However, we do not think that these potentials are compelling candidates, as they do not easily emerge from a UV-completion without extreme fine-tuning.   

In contrast, the simplest setup for hybrid inflation \cite{Linde:1993cn} with a slow-roll potential $V \sim V_0 + \frac{1}{2}m^2 \phi^2$, usually ruled out due to predicting a blue tilted spectrum, works well with warm inflation in the strong regime. As an example, we explore the inflationary dynamics for warm inflation in a hybrid setup in this section, where the inflaton field $\phi$ couples to a pure SU(3) gauge group, as described in Section \ref{AWI}.

The effective potential in hybrid inflation has two fields, one that acts as the inflaton $\phi$ and another the waterfall field $\sigma$ that stays constant during the inflationary period:
\begin{equation}
V(\phi, \sigma) = \frac{1}{4\lambda}\left(M^2 - \lambda \sigma^2 \right)^2 +\frac{1}{2}m^2 \phi^2 +\frac{1}{2} g^2 \phi^2 \sigma^2
\end{equation}
The squared mass of the waterfall field $\sigma$ is equal to $-M^2 + g^2 \phi^2$. While $\phi > \frac{M}{g}$, $\sigma$ only has one minimum at $\sigma = 0$. Inflation ends when $\phi$ reaches this threshold, which induces a first order phase-transition causing $\sigma$ to roll down to its minimum at $\sigma(\phi) = \frac{M_{\sigma}(\phi)}{\sqrt{\lambda}}$, with $M_\sigma(0) = M$. After the phase transition, $\phi$ rolls to the minimum of its effective potential  much faster than a Hubble time as long as:   
\begin{equation} \label{cond}
M^3 \ll \frac{\sqrt{\lambda} g m M^2_{\text{Pl}}}{Q}
\end{equation}
The waterfall field $\sigma$ rapidly starts oscillating after the phase transition as long as $M_{\sigma}(\phi) \gg H$. Under those conditions, inflation ends almost instantaneously. 

We can then describe the effective potential for the inflaton field $\phi$ during the time of inflation as: 
\begin{equation} \label{pot}
V_{\text{eff}}(\phi) = \frac{M^4}{\red{4}\lambda} + \frac{1}{2} m^2 \phi^2
\end{equation}
In the allowed parameter space outlined below, $\sigma$'s mass is larger than the temperature during inflation.  Thus, $\sigma$ does not thermalize and corrections to the thermal mass of $\phi$ turn out to be negligible. The observable amounts of e-folds occur as $\phi$ is approaching its critical value $\phi_c \equiv \frac{M}{g}$, which induces the phase transition. During this stage the constant term $\frac{M^4}{\lambda} \gg \frac{1}{2}m^2 \phi^2_c$ drives the expansion, effectively suppressing $\epsilon_V$. While $\phi$ is approaching its critical value it is sourcing a thermal bath via friction $\Upsilon$. The spectral index \eqref{ns} then simplifies to:
\begin{equation} \label{hybridns}
n_s -1 \approx -\frac{57}{7} \eta_V    
\end{equation}
with:
\begin{equation} \label{etaQ}
\eta_V = \frac{4\lambda m^2 M^2_{\text{Pl}}}{QM^4}     
\end{equation}
The spectral tilt fixes the following linear combination of parameters:
\begin{equation} \label{lc}
\frac{4\lambda m^2 M^2_{\text{Pl}}}{QM^4} \approx -\frac{7}{57} (n_s-1)    
\end{equation}
Assuming inequality Eqn.~\eqref{cond} is satisfied we can approximate $\phi_c \approx \phi_{\text{end}}$. Rewriting equation \eqref{NCMB} in the strong regime with $Q \gg 1$, and $\phi_{\text{CMB}} =(1+\Delta) \phi_c$, with $\Delta < 1$, we find:
\begin{equation} \label{intN}
N_{\text{CMB}}= \int^{\frac{M}{g}(1+\Delta)}_{\frac{M}{g}}\frac{1}{M^2_{\text{Pl}}} \frac{V}{V'} Q(\phi)d\phi
\end{equation}
 Using equation \eqref{papprox}, \eqref{Hprox}, and  \eqref{SS} we express $T$ and $Q$ in terms of $\phi$ during slow-roll, where $\rho_{R}=\frac{\pi^2}{30} g_{*} T^4 \equiv \tilde{g}_{*} T^4$, with $g_{*}$ denoting the relativistic degrees of freedom: 
\begin{equation} \label{SS-temperature}
T(\phi) \approx \left(\frac{f^2}{\kappa \alpha^5} \frac{\sqrt{3}  M_{\text{Pl}} {V'(\phi)}^2 }{\red{4} {\tilde{g_{*}}} \sqrt{V(\phi)}} \right)^{\frac{1}{7}}
\end{equation}

\begin{equation} \label{Qphis}
Q(\phi) \approx \left(\left(\frac{\kappa \alpha^5}{f^2}\right)^4 \frac{ M^{10}_{\text{Pl}}V(\phi)'^6}{576 {\tilde{g_{*}}}^{\red{3}}  V(\phi)^5}\right)^{\frac{1}{7}}
\end{equation}
Using equations \eqref{pot} and \eqref{Qphis} in equation \eqref{intN} and assuming $\frac{M}{g} \ll M_{\text{Pl}}$, we  obtain: 
\begin{equation} \label{Delta1}
N_{\text{CMB}} \approx \frac{\Delta}{\eta_V}    
\end{equation}
The number of observable e-folds, $N_{\text{CMB}} \approx 60$, then only impacts the transversed field range $\Delta$:
\begin{equation} \label{Delta}
\Delta \approx -\frac{7}{57} (n_s-1) N_{\text{CMB}}     
\end{equation}
Equation \eqref{Delta} determines $\Delta$ in terms of measured observables. Equation \eqref{lc} determines another linear combination of $\lambda, g, M, m, f, \Delta$ in terms of observables. The measured amplitude of the curvature power spectrum fixes one additional linear combination: 
\begin{equation}
\Delta^2_R(k) = A_s(k_{*})\left(\frac{k}{k_{*}}\right)^{n_s(k_{*})}
\end{equation}
with $A_s (\phi_{CMB}) \approx 2 \times 10^{-9}$ as measured by Planck at the pivot scale $k_{*}=0.05 \text{Mpc}^{-1}$ \cite{Aghanim:2018eyx}. Rewriting equation \eqref{spectrum} we find: 
\begin{equation}
A_s\left(\phi_{\text{CMB}}\right) \approx 8 \times 10^{-41} \left(\frac{\kappa \alpha^5}{f^2}\right)^{18}\left(\frac{{\sqrt{|n_s-1|}}^{81} m^{21} \phi^{51}_{\text{CMB}} }{{\sqrt{\tilde{g_{*}}}}^{55}} \right)^{\frac{1}{2}}     
\end{equation}
where $\phi_{\text{CMB}} = \frac{M}{g}(1+\Delta) \approx \frac{M}{g}$.


We have used the spectral index $n_s$, amount of observable e-folds $N_\text{CMB}$, and the amplitude of the power spectrum $A_s$, to constrain three of the parameters of the underlying model.  The friction ratio $Q$ depends on the ratio of the coupling $f$ to the field value of the inflaton $\sim \frac{M}{g}$ during inflation as:
\begin{equation}
Q \approx \red{120} \left(\frac{\Delta^2_R}{2 \times 10^{-9}}  \right)^{\frac{2}{21}}  \left(\frac{|n_s -1|}{0.035} \right)^{\frac{4}{7}}  \left(\frac{g_{*}}{17} \right)^{-\frac{4}{21}} \left(\frac{\kappa \alpha^5}{10^{-3}} \right)^{\frac{2}{7}}   \left(\frac{\frac{g f}{M}}{10^{-8}} \right)^{-\frac{4}{7}}
\end{equation}
Where we use a pure SU(3) with $g_{*}=17$ (two polarizations per eight gauge bosons plus one for the axion) and  gauge coupling $\alpha =0.1$ as our default values. The only tunable parameter beyond these is $\frac{g f}{M}$ which has to be $\lesssim 10^{-8}$ to place us in the strong regime ($Q \gg 1$), thus setting the upper bound $f \ll 10^{-8} (M/g)$ for these gauge group parameters. The typical Hubble scales and mass parameters in our model are thus: 
\begin{equation} \label{Hubble}
H \approx \red{2.5 \times 10^{-16}} \left(\frac{\Delta^2_R}{2 \times 10^{-9}} \right)^{\frac{1}{21}}  \left(\frac{|n_s -1|}{0.035} \right)^{-\frac{19}{7}} \left(\frac{g_{*}}{17} \right)^{\frac{59}{42}} \left(\frac{\kappa \alpha^5}{10^{-3}} \right)^{-\frac{13}{7}}   \left(\frac{\frac{gf}{M}}{10^{-8}} \right)^{\frac{\red{26}}{7}} \frac{M}{g}
\end{equation}

\begin{equation}
m  \approx \red{3 \times} 10^{-16} \left(\frac{\Delta^2_R}{2 \times 10^{-9}} \right)^{\frac{2}{21}} \left(\frac{|n_s -1|}{0.035} \right)^{-\frac{12}{7}}    \left(\frac{g_{*}}{17} \right)^{\frac{55}{42}} \left(\frac{\kappa \alpha^5}{10^{-3}} \right)^{-\frac{12}{7}}   \left(\frac{\frac{gf}{M}}{10^{-8}} \right)^{\frac{24}{7}} \frac{M}{g}
\end{equation}
Note that $m$ can be larger than $H$, without violating slow-roll due to the dominant friction coming from $\Upsilon\gg H$. Typical temperatures during expansion are given by: 
\begin{equation}
T  \approx   \red{2 \times 10^{-9}} \left(\frac{\Delta^2_R}{2 \times 10^{-9}} \right)^{\frac{1}{21}} \left(\frac{|n_s -1|}{0.035} \right)^{-\frac{12}{7}}    \left(\frac{g_{*}}{17} \right)^{\frac{17}{42}} \left(\frac{\kappa \alpha^5}{10^{-3}} \right)^{-\frac{6}{7}}   \left(\frac{\frac{gf}{M}}{10^{-8}} \right)^{\frac{12}{7}} \frac{M}{g}
\end{equation}

Demanding that condition \eqref{cond} is satisfied such that inflaton quickly rolls to its minimum after the phase transition imposes an upper limit on  $\frac{M}{g}$:
\begin{equation} \label{cond2}
 \frac{M}{g} \ll 3 \times 10^{-3} \left(\frac{\Delta^2_R}{2 \times 10^{-9}} \right)^{-\frac{1}{21}} \left(\frac{|n_s -1|}{0.035} \right)^{\frac{3}{14}}  \left(\frac{g_{*}}{17} \right)^{\frac{2}{21}} \left(\frac{\kappa \alpha^5}{10^{-3}} \right)^{-\frac{1}{7}}   \left(\frac{\frac{gf}{M}}{10^{-8}} \right)^{\frac{2}{7}} M_{\text{Pl}}   
\end{equation}
The above condition demands that the maximum allowed value for $\frac{M}{g}$ is roughly $10^{14}\,$GeV. This value sets an upper limit for the possible temperatures of $T < 5 \times 10^4\,$GeV and Hubble scales of $H < 10^{-3} \,$GeV. The discussed observables degenerately depend on combinations of $M$, $\lambda$ and $g$. Requiring the quantum corrections to our masses be naturally small also imposes constraints that break the degeneracy:   
\begin{equation} \label{n1}
\frac{\lambda^2 \Lambda^2}{16 \pi^2} < M^2 
\end{equation}
\begin{equation} \label{n2}
\frac{g^2 \Lambda^2}{16 \pi^2} <m^2
\end{equation}
where $\Lambda$ is the cutoff of the theory. The couplings $g$ and $\lambda$ need to satisfy conditions \eqref{n1} and \eqref{n2}. Additionally, the condition that the $\phi$ potential is negligible compared to the vacuum energy during inflation requires: 
\begin{equation} \label{c3}
\frac{\lambda  m^2}{g^2} \ll M^2     
\end{equation}
Assuming a minimum value of the cutoff $\Lambda = 4 \pi M$, saturating equation \eqref{n2} and $\frac{M}{g} =10^{14}\,$GeV, \red{\sout{and satisfying \eqref{c3} by two orders of magnitude,}} we get the following sample values for the couplings and mass parameters: $g =10^{-8}$, $\red{\lambda = 2 \times 10^{-11}}$, $M = 10^{6}\,$GeV, $m = \red{0.03}$ GeV. 


\subsection{Reheating}
At reheating, we assume that we have an abundance of $\sigma$ particles at some early time before big bang nucleosynthesis, which make up a dominant part of the energy density in the early universe. There are many possible ways in which $\sigma$ can couple to the standard model and produce an early quark gluon plasma. Here we outline a simple example where we couple to standard model hypercharge:
\begin{equation} \label{op}
\mathcal{L}_{\text{reheat}} = \frac{{g'}^2}{64 \pi^2}\frac{\sigma}{f_{B}}\tilde{B}^{\mu\nu} B_{\mu\nu} 
\end{equation}    
where $g'$ denotes the standard model hypercharge gauge coupling. Typical values of  the coupling between the waterfall and inflaton fields in our model ($g$) are quite small, which is why $\sigma$ decays dominantly via operator \eqref{op}, even for large values of $f_B$. 
We can estimate the decay rate into standard model particles by: 
\begin{equation}
\Gamma_{\sigma \to \text{SMSM}} = \frac{{g'}^4 M^3}{16384 \pi^5 f^2_B}
\end{equation}
This rate needs to be large enough such that an abundance of $\sigma$ particles has decayed into a quark gluon plasma before the universe cools down to  big bang nucleosynthesis temperatures, where the earliest cosmological precision constraints exist. We estimate the Hubble rate as $H_{\text{BBN}} \approx \frac{\sqrt{\frac{ g_{*}\pi^2 }{30}}T_{\text{BBN}}^2}{\sqrt{3}M_{\text{Pl}}}$ and demanding that $\Gamma_{\sigma \to \text{SMSM}} > H_{\text{BBN}}$ we find that $f_B$ can go all the way up to the GUT scale: 
\begin{equation}
f_B <   10^{16} \, \text{GeV} \left( \frac{M}{10^6\, \text{GeV}} \right)^{\frac{3}{2}}   \left( \frac{T_{\text{BBN}}}{10\, \text{MeV}}\right)^{-1}      
\end{equation}
\\
At the end of section \ref{Iitm} we briefly discuss the upper limits of masses, temperatures and Hubble scales. Here we discuss the lower limits of our parameters. Since the waterfall field $\sigma$ couples directly to the standard model there exist cooling bounds from supernovae as well as detection constraints from high-energy colliders. Avoiding these, we conservatively set $f_B > 1 \,$TeV and $M > 10\,$GeV as the lower limits of our parameter space which fixes $m > 10^{-7}\,$GeV, $H > 10^{-8}\,$GeV and $T > 0.5\,$GeV, where $T$ is the temperature during slow-roll maintained by the pure Yang-Mills radiation. These parameters easily still satisfy the cosmological constraints that reheating happens efficiently before BBN. Summarizing our available parameter space we find these:
\begin{align}
10^{-8}\,\text{GeV} &< H \, \red{\lesssim  10^{-2}}\,\text{GeV}\\
10^{-7}\,\text{GeV} &< m   \, \, \lesssim 10^{-2}\,\text{GeV}\\
10\,\text{GeV} &< M  \lesssim 10^{6}\,\text{GeV}\\
0.5\,\text{GeV} &< T \, \red{\lesssim  10^{5}}\,\text{GeV}\\
0.5\,\text{GeV} &< f  \lesssim \red{ 10^{6}}\,\text{GeV}
\end{align}
are the maximally allowed ranges for each parameter, though of course there are restrictions on the combination of the five parameters (e.g.~the requirement of decay before BBN and the validity of the effective field theory).
The question remains whether the inflaton coupling to a standard model gauge group itself (e.g.~QCD) can give rise to a thermal bath sourcing friction during inflation. In that scenario, a quark gluon plasma is already present during the expansion of the universe and reheating becomes trivial. However, currently detailed calculations of the friction coefficient exist only for pure Yang-Mills theory. The presence of light fermions may non-trivially alter the parametric dependence\footnote{This concern was pointed out to us by members of the theory group at the University of Maryland.  We thank them and Guy Moore for extensive discussions on this topic.}, in which case a separate analysis is necessary to determine whether this compelling simplification is viable. We leave that analysis and the calculation of the friction in the presence of light fermions to future work.

\section{Conclusions}\label{Conclusions}

If the inflaton has any non-gravitational coupling to other fields it will generically produce a background thermal bath during inflation.  A natural choice is an axion-like (CP-odd) coupling which can generate significant thermal friction from non-perturbative effects for the inflaton without a corresponding backreaction on the inflaton potential, thus avoiding the problems with other warm inflation models.  Once the inflaton has any such strong enough coupling, a thermal bath will necessarily be produced during inflation independent of initial conditions.

We have presented a complete model of warm inflation which correctly reproduces cosmological data on initial density perturbations and predicts a negligible tensor-to-scalar ratio $r$ and potentially measureable non-gaussianities.  The model only requires the inflaton to have an axion-like coupling to a non-Abelian group, and we use known results for couplings to pure Yang-Mills.
An even simpler model may be possible where the inflaton couples directly to the standard model (such as to QCD), but a full thermal field theory calculation of the friction in this case (specifically with light quarks) has not yet been done.
We show, as an example,  that the  temperature dependence of the friction due to our coupling allows 
hybrid inflation to have a red-tilted spectrum (rather than blue-tilted as in cold inflation), and thus can easily fit the current data. 

\section*{Acknowledgments}
We are grateful to Mikko Laine and Guy Moore for extensive conversations about the damping effects of sphalerons in different contexts.  We would also like to thank Gauraw Gosmani, Marc Kamionkowski, Alvise Raccanelli, Surjeet Rajendran, Tristan Smith,   Tommi Tenkanen and the University of Maryland theory group for helpful discussions. \red{We thank Mark Vincent Guevarra for pointing out several minor mistakes in equations in Sec.4 in earlier versions.} We acknowledge the support of NSF Grant PHY-1818899,
NSF Grant PHY-1720397, DOE Grant DE-SC0012012,  the Heising-Simons Foundation Grants 2015-037 and 2018-0765, DOE HEP QuantISED award \#100495, and the Gordon and Betty Moore Foundation Grant GBMF7946.

\appendix

\section{The Weak Regime}
\label{sec: Weak regime appendix}
Due to the calculations of the friction coefficient breaking down in the weak regime we have focused on exploring the strong regime ($Q \gg 1$) in detail in this paper. In this appendix we summarize the relevant dynamics in the weak regime.
It turns out that the only viable models of warm inflation we could find in the weak regime require parameters which move the thermal friction beyond the regime of validity of the thermal field theory calculations which have been done.
Thus in this section we will simply assume that the friction coefficient still scales as $\Upsilon \sim \kappa \alpha^5 \frac{T^3}{f^2}$, and discuss warm inflation in this case.  But we will find that in fact we are ultimately pushed to a regime of parameters where this formula is not known to be valid.  So it is in fact possible that a weak warm inflation model would work -- even for a simple inflaton potential $m^2 \phi^2$ -- but we cannot know that from the thermal field theory calculations that have been done to date.

In the weak regime of warm inflation ($Q \ll 1$) the dominant friction in the inflaton's equation of motion is still due to the Hubble expansion rather than particle production friction. However, the presence of a thermal bath can still change the power spectrum and effectively surpress the scalar-to-tensor ratio. Unlike in the strong regime the temperature dependent friction coefficient does not give rise to a growing mode as the coupling between the radiation and the inflaton can be neglected. The curvature power spectrum  and the scalar to tensor ratio in this regime can then be described by \cite{Berera:1995wh, Bartrum:2013fia, Visinelli:2014qla}, where all quantities are evaluated at horizon crossing:
\begin{equation} \label{spectrumweak}
\Delta^2_R = \frac{1}{4 \pi^2}\frac{H^4}{\dot{\phi}^2} \left(1 + 2n +2\pi Q\frac{T}{H} \right)
\end{equation}
\begin{equation}
r=\frac{16 \epsilon_V}{\left(1 + 2n +2\pi Q\frac{T}{H} \right)}    
\end{equation}
Here $n$ denotes the distribution of inflaton particles. If interactions between the inflaton particles and the thermal bath are sufficiently fast for them to be thermalized then they approach a Bose-Einstein distribution, which at horizon crossing is given by $n_{\text{BE}} = \left(e^{\frac{H}{T}}-1 \right)^{-1}$. Whether thermalization is fast enough is model dependent. The interaction rate for the axion-inflaton with the gauge boson radiation, $\Gamma_{\phi g}$, can be roughly approximated as $\Gamma_{\phi g} \approx \alpha^3  \frac{T^3}{32 \pi f^2} = \frac{\Upsilon}{32 \pi \kappa \alpha^2}$. The inflaton is thermalized ($\Gamma_{\phi g} >H$), when $\frac{3Q}{32 \pi \kappa \alpha^2} > 1$, which is satisfied for $\alpha \lesssim 10^{-2} \sqrt{Q}$. Thus, whether thermalization occurs depends on the gauge coupling of the YM-group itself. For a temperature dependence of the friction $\Upsilon \propto T^3$, we can derive the spectral index in the weak regime using equations \eqref{dlnH}, \eqref{dlnphid}, \eqref{dlnU}, \eqref{si} and: 
\begin{equation}
\lim_{Q \ll 1}\frac{d {\ln T}}{dN} = \left(3 \epsilon_V - 2 \eta_V \right)
\end{equation}
finding: 
\begin{equation} \label{nsw}
n_s-1 = \frac{1}{1+2n +\frac{2 \pi Q T}{H}} \left(2 \eta_V - 6 \epsilon_V  \right)+ \frac{\frac{2 \pi Q T}{H}}{1+ 2n+\frac{2 \pi Q T}{H}} \left( 8 \epsilon_V - 6 \eta_V \right) +\frac{2n}{1+2n +\frac{2 \pi Q T}{H}}\left(-2\epsilon_V \right)
\end{equation}
If the inflaton is not thermalized and $\frac{2 \pi Q T}{H} \ll 1$ we recover the regular cold inflation result. The size of this parameter determines whether we are in a regime in which thermal fluctuations dominate over quantum effects. When quantum effects dominate the spectral index obtains a higher order correction, which is negligible. However, when thermal fluctuations dominate we again obtain a spectral index that can only be red-tilted for potentials where $\eta_V$ dominate, again demanding a fine-tuning of single field potentials, similarly to the strong case. If the inflaton is fully thermalized the third term in equation \eqref{nsw} dominates as $n \sim \frac{T}{H}$ and $Q \ll 1$. However, for single field potentials the predicted spectral index $n_s$ for about $50$ to $60$ e-folds lies outside of the two sigma region. There does exist a transition region where the inflaton is not fully thermalized for $n \sim \frac{2 \pi Q T}{H} < 1$, where the observed spectral tilt can be reproduced in the weak regime. However, in this transition region non-gaussianity constraints become important \cite{Bastero-Gil:2014raa}. For detailed non-gaussianity predictions in the weak regime in the presence of a friction that scales as $\Upsilon \propto T^3$, see \cite{Bastero-Gil:2014raa}.

The weak warm inflation formulas above have only been calculated in the regime where $Q \ll 1$ (for a temperature-dependent friction coefficient).  Additionally, being conservative we are only certain we can trust the thermal field theory calculations when $\alpha^2 T > H$ and $\alpha^2 T > m$ (where $m$ is the mass of the inflaton).  Taking the combination of all these constraints on the validity of the calculations that have been done, we find no region of parameter space that can fit the observations (the values of $n_s$, $r$, number of e-folds and the size of the perturbations).  So we are unable to make an observationally viable weak warm inflation model.  However it is possible that if the calculations for warm inflation were extended to include a region of $Q \sim 1$ (for our temperature-dependent friction) one could find a viable inflation model.  Or similarly it is possible that if the thermal field theory calculations were valid beyond $\alpha^2 T > H$ and $\alpha^2 T > m$ then one could find a viable weak warm inflation model.  We leave this for future work.

\bibliography{WIbib}

\end{document}